\documentclass[preprint,12pt, a4paper]{elsarticle}

\usepackage{xcolor}
\usepackage{listings}

\definecolor{codebg}{RGB}{245, 245, 245}
\definecolor{stringcolor}{RGB}{195, 75, 75}
\definecolor{keywordcolor}{RGB}{0, 76, 153}
\definecolor{commentcolor}{RGB}{85, 107, 47}
\definecolor{numbercolor}{RGB}{138, 43, 226}

\lstdefinestyle{customStyle}{
    language=Python,
    backgroundcolor=\color{codebg},
    keywordstyle=\color{keywordcolor}\bfseries,
    commentstyle=\color{commentcolor},
    stringstyle=\color{stringcolor},
    numberstyle=\color{numbercolor},
    basicstyle=\ttfamily\footnotesize\ttfamily,
    showstringspaces=false,
    showspaces=false,
    breaklines=true,
    frame=leftline,
    rulecolor=\color{gray},
    numbers=none,
    xleftmargin=2pt,
    xrightmargin=2pt,
    tabsize=2,
    captionpos=b
}

\makeatletter
\def\els@aparagraph[#1]#2{\elsparagraph[#1]{#2}}
\def\els@bparagraph#1{\elsparagraph*{#1}}
\makeatother

\usepackage{amsmath} 
\usepackage{booktabs}
\usepackage{comment}
\usepackage{float}
\usepackage{hyperref}
\usepackage{listings}
\usepackage{amsfonts}
\usepackage{paralist}
\usepackage{subcaption}
\begin{document}

\makeatletter
\def\ps@pprintTitle{%
\let\@oddhead\@empty
\let\@evenhead\@empty
\let\@oddfoot\@empty
\let\@evenfoot\@empty}
\makeatother

    \renewcommand{\labelenumii}{\arabic{enumi}.\arabic{enumii}}

    \begin{frontmatter}
        \title{RouteRL: Multi-agent reinforcement learning framework for urban route choice with autonomous vehicles}
        \author[phd]{Ahmet Onur Akman*}
\author[phd]{Anastasia Psarou*}
\author[mem]{Łukasz Gorczyca}
\author[phd]{Zoltán György Varga}
\author[mem]{Grzegorz Jamróz}
\author[mem]{Rafał Kucharski}

\address[phd]{Doctoral School of Exact and Natural Sciences, Jagiellonian University, Kraków, Poland}
\address[mem]{Faculty of Mathematics and Computer Science, Jagiellonian University, Kraków, Poland}
        \begin{abstract}
    RouteRL is a novel framework that integrates multi-agent reinforcement learning (MARL) with a microscopic traffic simulation, facilitating the testing and development of efficient route choice strategies for autonomous vehicles (AVs). The proposed framework simulates the daily route choices of driver agents in a city, including two types: human drivers, emulated using behavioral route choice models, and AVs, modeled as MARL agents optimizing their policies for a predefined objective. RouteRL aims to advance research in MARL, transport modeling, and human-AI interaction for transportation applications. This study presents a technical report on RouteRL, outlines its potential research contributions, and showcases its impact via illustrative examples.
\end{abstract}

    \end{frontmatter}
    
    \section*{Metadata}
\label{sec:metadata}

    \begin{table}[ht]
\begin{tabular}{|l|p{6.5cm}|p{6.5cm}|}
\hline
\textbf{Nr.} & \textbf{Code metadata description} & \textbf{Please fill in this column} \\
\hline
C1 & Current code version & v1.0.0\\
\hline
C2 & Permanent link to code/repository used for this code version & \url{https://github.com/COeXISTENCE-PROJECT/RouteRL/tree/v1.0.0} \\
\hline
C3  & Permanent link to Reproducible Capsule & \url{https://codeocean.com/capsule/6105686/tree/v1}\\
\hline
C4 & Legal Code License   & MIT License \\
\hline
C5 & Code versioning system used & git \\
\hline
C6 & Software code languages, tools, and services used & Python, SUMO, OpenStreetMap, PettingZoo, Pytorch, TorchRL, CUDA, Sphinx, Jupyter Notebook\\
\hline
C7 & Compilation requirements, operating environments \& dependencies & Python ($\ge$ 3.8, $\le$ 3.13), OS independent\\
\hline
C8 & If available Link to developer documentation/manual & \url{https://coexistence-project.github.io/RouteRL/} \\
\hline
C9 & Support email for questions & coexistence@uj.edu.pl
\\
\hline
\end{tabular}
\caption{Code metadata}
\label{codeMetadata} 
\end{table}
    \section{Motivation and significance}\
\label{sec:motivation}

Autonomous vehicles (AVs) will soon be ready to enter our cities and make routing decisions \cite{av_predictions, urban_future}. Using computational power, access to data, and real-time information, AVs will be able to develop efficient route choice strategies to reach their destinations. Presumably, their route choices will deviate from those made by human drivers, currently using our urban networks. Yet the routing strategies to be adopted by AVs as well as their impact on the urban traffic systems and other users (human drivers) remain unknown. 

To better understand this newly arising class of problems, we introduce the RouteRL framework. RouteRL reproduces the traffic flow on urban networks (simulated with SUMO \cite{SUMO}) shared by human drivers and (connected) autonomous vehicles. Humans, while selecting routes, maximize perceived utility (reward), represented with state-of-the-art behavioral route-choice models. Meanwhile, AVs employ various multi-agent reinforcement learning (MARL) algorithms to develop efficient route-choice strategies. Every day, each agent selects subjectively optimal routes on consecutive days of the learning process. RouteRL simulates how experienced humans and trained MARL agents traverse urban networks, and eventually, outputs a series of system performance indicators.

\begin{figure}[ht]
                \centering
                \includegraphics[width=0.99\textwidth]{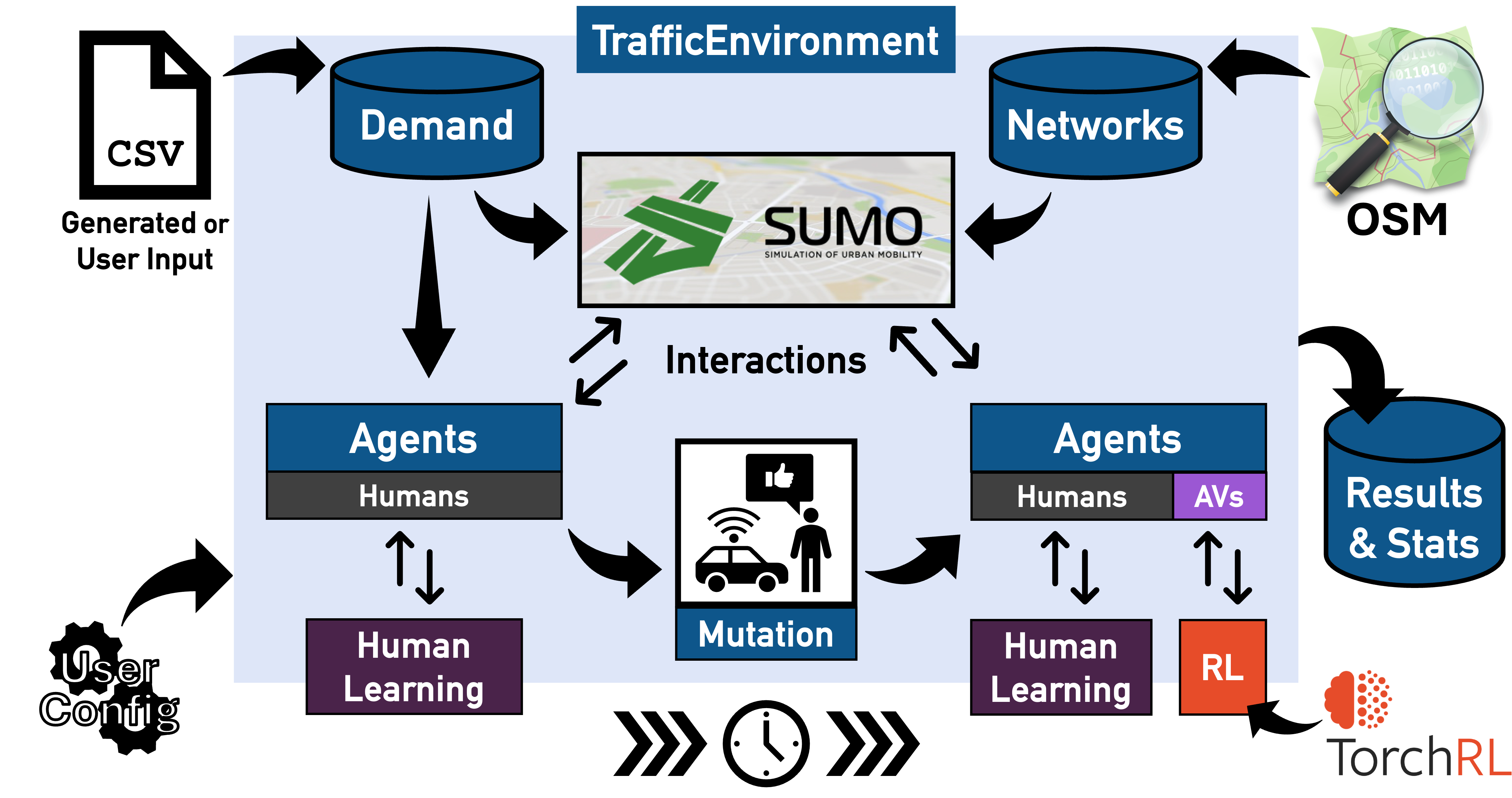}
                \caption{
                    \textbf{The software at a glance:} 
                    RouteRL can simulate any OSM network on which a set of agents demands to reach their destinations. SUMO traffic microsimulator simulates how vehicles traverse the network. In a sequence of days (episodes), agents iteratively learn how to optimize their routing decisions (arrive faster). Humans use behavioral models, and AVs use policies trained with state-of-the-art MARL algorithms implemented via TorchRL.
                }
                \label{fig:routerl_flow}
            \end{figure}

RouteRL allows for a broad experimental schema to investigate the impact of the future of AVs on urban networks. The pipeline of a typical RouteRL experiment, like the ones presented in the paper, is depicted in Figure \ref{fig:routerl_flow}. Given a road network (from OpenStreetMap (OSM) \cite{OpenStreetMap}) and a demand pattern (agents' origins, destinations, and departure times), RouteRL first computes the day-to-day route-choice process of human agents that eventually achieve stabilization (though not necessarily reaching a Nash User Equilibrium \cite{wardrop_road_1952}). Then, a predefined portion of human drivers \emph{mutate} to AVs. Notably, unlike most recent research, here AVs have the same traffic properties (acceleration rate, gap acceptance, etc.) as human drivers - which allows disentangling effects of optimized route choice from improved traffic performance. AVs may use a variety of MARL algorithms either from TorchRL \cite{torchrl} or from custom implementations to train their policies (individually or collaboratively) and maximize their rewards (representing behaviors ranging from selfish to social or malicious).
Specifically, each AV every day selects from the finite action space (possible routes) the optimal (in terms of expected reward) route to reach its destination under the current state observation. This choice is either individual or managed by some central fleet manager. Agents do not change the action within the day and follow the chosen route up to the destination, having a chance to switch routes only on the following day (episode). Depending on the scenario, humans may have a chance to adapt their decisions, follow the previous choices, or transition into AVs,~each possibly triggering new day-to-day dynamics. RouteRL records travel times, rewards, and space-time trajectories of individual vehicles, aggregating them into key performance indicators (KPIs) to assess the impact on system performance and different user groups.

Through such experiments, RouteRL allows investigating relevant questions arising with the introduction of AVs into our cities, such as:
\begin{compactitem}
    \item In the Cologne network, if 40\% of drivers switch to AVs and collaboratively maximize human travel time using the QMIX algorithm \cite{qmix} what will be the impact on total travel times (Figure \ref{fig:action_shifts_travel_times})? 
    \item In the Ingolstadt network, if 20\% of drivers transition to AVs and individually maximize their reward, what is the most efficient algorithm (Figure \ref{fig:mean_rewards}a)?
    \item In the Manhattan network, if 12.5\% of users become altruistic  AVs (aim to minimize travel time of all the drivers in the network), is it more advantageous to remain a human driver or switch (Figure \ref{fig:cav_advantage})?
\end{compactitem}

It is fundamental to address this class of problems before allowing AVs to integrate into urban networks. Upfront experiments with RouteRL are instrumental in informing: 
\begin{compactitem}
    \item policymakers (on potential consequences for system performance),
    \item transportation engineers (on impact on urban traffic), and
    \item the machine learning community (on the efficiency of (MA)RL algorithms for route choice).
\end{compactitem}
    \section{Software description}
\label{sec:software_description}
    
    Finding efficient route choice strategies for a group of vehicles (like a fleet of AVs) in a mixed traffic system is a dynamic optimization problem. RouteRL models this as a multi-agent decision-making task and integrates it with MARL.
     RouteRL main class (\texttt{TrafficEnvironment}) adheres to PettingZoo Agent Environment Cycle (AEC) API, providing a standardized interface for MARL. Experimenting using RouteRL is similar to working with any standard PettingZoo environment (as demonstrated with the code snippet below). The training pipeline involves an environment instantiation and training policies by interacting with it. Moreover, the MARL training procedure fully integrates with TorchRL, enabling users to leverage optimized and actively maintained implementations of state-of-the-art MARL algorithms, while preserving customizability.
    
\begin{figure}[ht]
        \centering
        \includegraphics[width=0.9\textwidth]{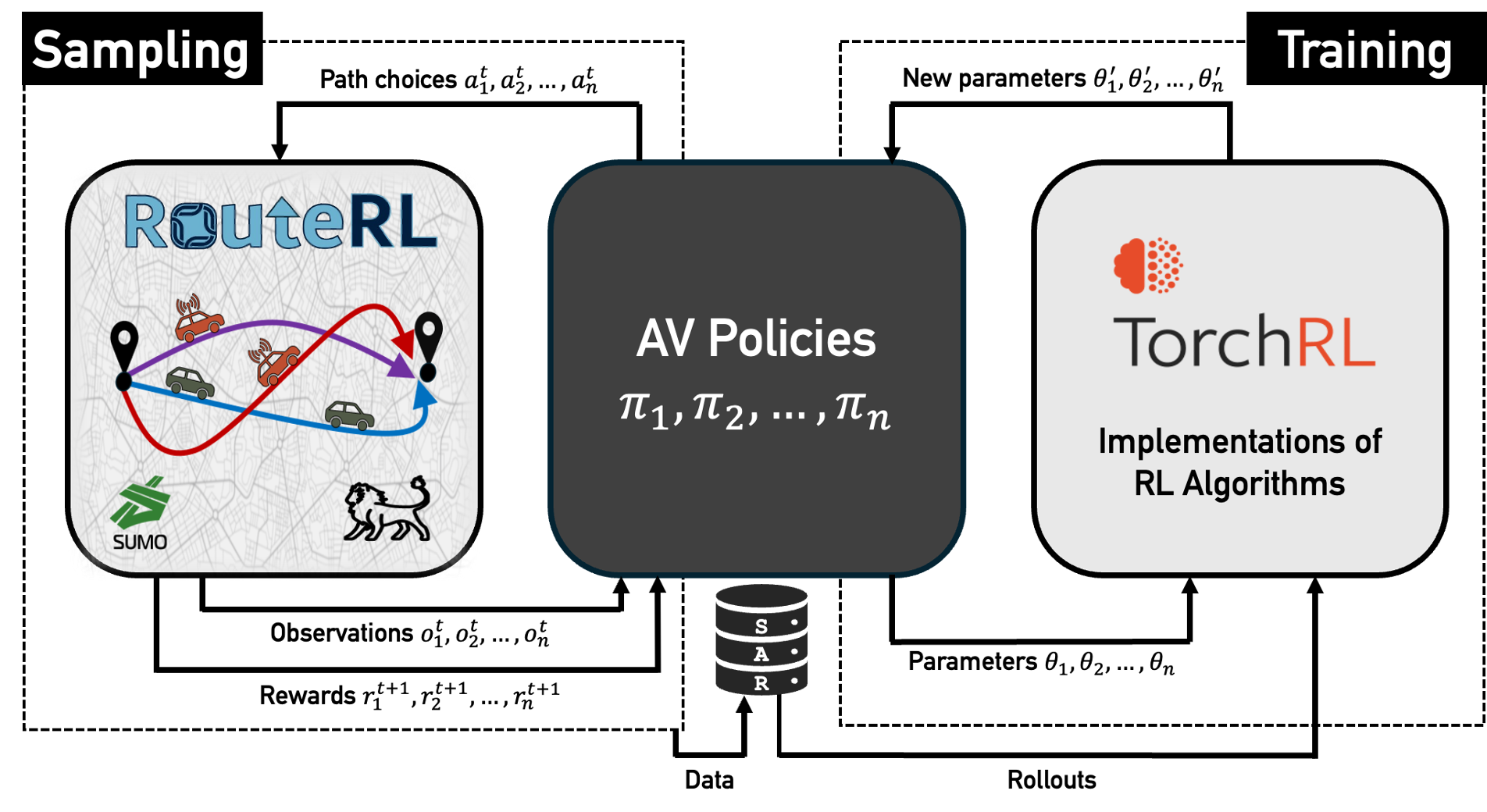}
        \caption{
        \textbf{A typical multi-agent training pipeline using RouteRL}: The training procedure alternates between two stages: (i) \textit{Sampling} (left), where traffic microsimulator (SUMO) returns rewards resulting from given route choices made by all agents, and (ii) \textit{Training} (right), where the collected experiences are used to train AV policies via TorchRL’s MARL algorithm implementations. Policy parameters are iteratively updated and fed back into the sampling loop.
        }
        \label{fig:training_flow}
    \end{figure}

    By integrating widespread frameworks (such as PettingZoo, TorchRL, and TraCI) and standard input/output formats (keyword arguments, \texttt{OSM} files, \texttt{CSV} files), RouteRL ensures interoperability across frameworks and streamlines experiment setup and data analysis. Figure \ref{fig:training_flow} illustrates a high-level workflow for conducting MARL training using RouteRL.
    
    \subsection{Software architecture}
    \label{sec:architecture}

        RouteRL is a Python open-source package built with a modular design, where each component serves distinct but interconnected purposes, and their modules decompose responsibilities into isolated units. It has three subpackages: \texttt{environment} (contains components related to the environment's workflow), \texttt{human\textunderscore learning} (includes human route choice and learning models), and \texttt{services} (I/O operations and data visualization). The UML class diagram in \ref{sec:class_diagram} illustrates the class hierarchy of RouteRL, while the sequence diagram in \ref{sec:seq_diagram} depicts the user interaction.
        Figure \ref{fig:routerl_flow} illustrates the workflow of \texttt{RouteRL}, showing the interactions during a typical experiment run and the resulting input-output flow. 
        
        The \texttt{environment.TrafficEnvironment} class functions as the central orchestrator and a MARL environment. It provides users with an intuitive interface for key functionalities, such as: 
        \begin{compactitem}
            \item \textbf{Agent-environment interaction}: Functioning as a MARL environment and handling agent interactions for training.
            \item \textbf{Simulation control}: Through a \texttt{SumoSimulator} object, handling the simulation control, executing agent actions in SUMO, and retrieving episode statistics for training.
            \item \textbf{Demand generation}: Generating travel demand data (origin, destination, and departure of each agent) synthetically or importing from user-provided \texttt{CSV} files.
            \item \textbf{Mutation}: Handling how human agents transition into AVs and initiating MARL training loop.
            \item \textbf{Data recording and visualization}: With a \texttt{Recorder} object storing episode data during training, used (after the training) by a \texttt{Plotter} object to generate insightful summaries with data visualizations (as depicted in \ref{sec:plots}).  
        \end{compactitem}

    \subsection{Software functionalities}
    Key RouteRL functionality is to run experiments like the one depicted in Figure \ref{fig:routerl_flow} with a unified workflow, flexibility in parameterization, and the reproducibility needed for scientific rigor. Users can configure custom scenarios via keyword arguments during environment initialization (as detailed in our online documentation). Possible customizations include specifying the network, demand pattern, human route choice model, mutation day, the share of AVs, and their behaviors. Moreover, users can implement and experiment with their custom human behavioral models and custom MARL implementations. 
    At the end of a training, RouteRL outputs raw and aggregated results.

        \subsubsection{Modeling the decision process}
        \label{sec:interactions}
            From an individual agent’s perspective, daily route choice is a single-step decision: each agent makes a route choice, and the episode ends upon their arrivals at their destinations. The environment is built on the PettingZoo AEC API, therefore the agent interactions are sequential (in order of departure time). This allows agents to observe the most recent traffic conditions before making route choice decisions.
            \emph{State transitions} are governed by SUMO, which is a state-of-the-art, microscopic, agent-based traffic simulator. Each vehicle navigates the road network according to the Intelligent Driver Model (IDM) \cite{Treiber_2000}. In each episode, SUMO simulates traffic flow dynamics based on individual route choices and computes agents' travel times.
            
            For each agent, the \emph{action space} includes discrete route options, created with a dedicated path generator tool JanuX \cite{janux}. 
            The \emph{observation} includes past route choices from earlier timesteps within the same episode. Users can also incorporate departure times into observations, which is useful for centralized or shared-parameter training.
            The AV \emph{reward signal} is designed to represent a selected AV behavior. RouteRL provides users with a selection of AV route choice behaviors to experiment with (see Table \ref{tab:behaviors}), defined through different AV reward formulations (adopted from \cite{akman_impact, jamroz}). 
                    \begin{table}[ht]
        \centering
        \caption{AV behaviors and corresponding objectives}
            \begin{tabular}{c|l}
            \toprule
            \textbf{AV Behavior} & \textbf{Objective} \\ 
            \midrule
            Altruistic    & Minimize delay for everyone                             \\
            Collaborative & Minimize delay for oneself and one's own group          \\
            Competitive   & Minimize self-delay and maximize for the other group        \\
            Malicious     & Maximize delay for the other group                           \\
            Selfish       & Minimize delay for oneself                               \\
            Social        & Minimize delay for oneself and everyone                 \\ 
            \bottomrule
            \end{tabular}
        \label{tab:behaviors}
        \end{table}

        \subsubsection{Human learning}
        \label{sec:humans}
            The \texttt{human\textunderscore learning} package includes three state-of-the-art discrete route choice models: \texttt{Gawron} \cite{gawron}, \texttt{CULO} \cite{culo}, and \texttt{WeightedAverage} \cite{cascetta}. These models, popular within the transportation community, emulate human agents as \textit{utility maximizers}, where individual utilities are influenced by individual characteristics \cite{cascetta}—unlike MARL algorithms that primarily focus on cost minimization.

        \subsubsection{Traffic networks}
        \label{sec:networks}
            RouteRL includes eleven traffic networks varying in size and characteristics, including: "arterial", "cologne", "grid", "ingolstadt" (from RESCO benchmark \cite{resco}), "grid6", "square" (from SUMO's repository \cite{sumo_repo}), "ortuzar", "nguyen" (from SUMO-RL \cite{sumorl}), "manhattan", "two\_route\_yield" and "csomor" networks (first included in RouteRL). Network layouts are shown in our online documentation.

        \subsubsection{Reproducibility}
        \label{sec:reproducibility}
            Providing a fixed random seed to \texttt{TrafficEnvironment} ensures complete experiment reproducibility. The seed is used to initialize all stochastic events within the environment, including simulation dynamics (SUMO), demand generation, route generation (JanuX), and human decision-making. This facilitates reliability for benchmarking, debugging, and comparative analysis, which are crucial for scientific impact.
      
    \subsection{Sample code snippets analysis}
    \label{sec:snippets}
    
        Listing \ref{code:snippet} demonstrates how a user interacts with RouteRL to conduct training on a standard MARL algorithm implementation via TorchRL.
    
\begin{lstlisting}[
style=customStyle, 
caption={\textbf{Standard RouteRL usage}: In this simplified code, the user initializes the environment, imports the road network and demand pattern (agents) (1), connects with SUMO (2), simulates learning of human drivers (3) which, after mutation (4), interact with the environment and collect data (5) to train MARL policies (6), which are finally tested and reported (7).}, 
label={code:snippet}]
# (1) initialize the traffic environment
env = TrafficEnvironment(seed=42, **env_params)

# (2) start the connection with SUMO
env.start()

# (3) human learning 
for episode in range(human_learning_episodes):
    env.step()

# (4) some human agents transition into AV agents
env.mutation()

# (5) collects experience by running the policy in the environment (TorchRL)
collector = SyncDataCollector(env, policy, ...)

# (6) training of the AVs
for tensordict_data in collector:        
    # update the policies of the learning agents
    for _ in range(num_epochs):
      subdata = replay_buffer.sample()
      loss_vals = loss_module(subdata)
      optimizer.step()
    collector.update_policy_weights_()

# (7) testing phase and storing results
policy.eval()
for episode in range(100):
    env.rollout(len(env.machine_agents), policy=policy)
env.plot_results() # plot the results
env.stop_simulation() # close the connection with SUMO
\end{lstlisting}
    \section{Illustrative examples}
\label{sec:examples}

\begin{figure}[ht]
    \centering
    \begin{subfigure}[b]{0.45\textwidth}
        \centering
        \includegraphics[width=\textwidth]{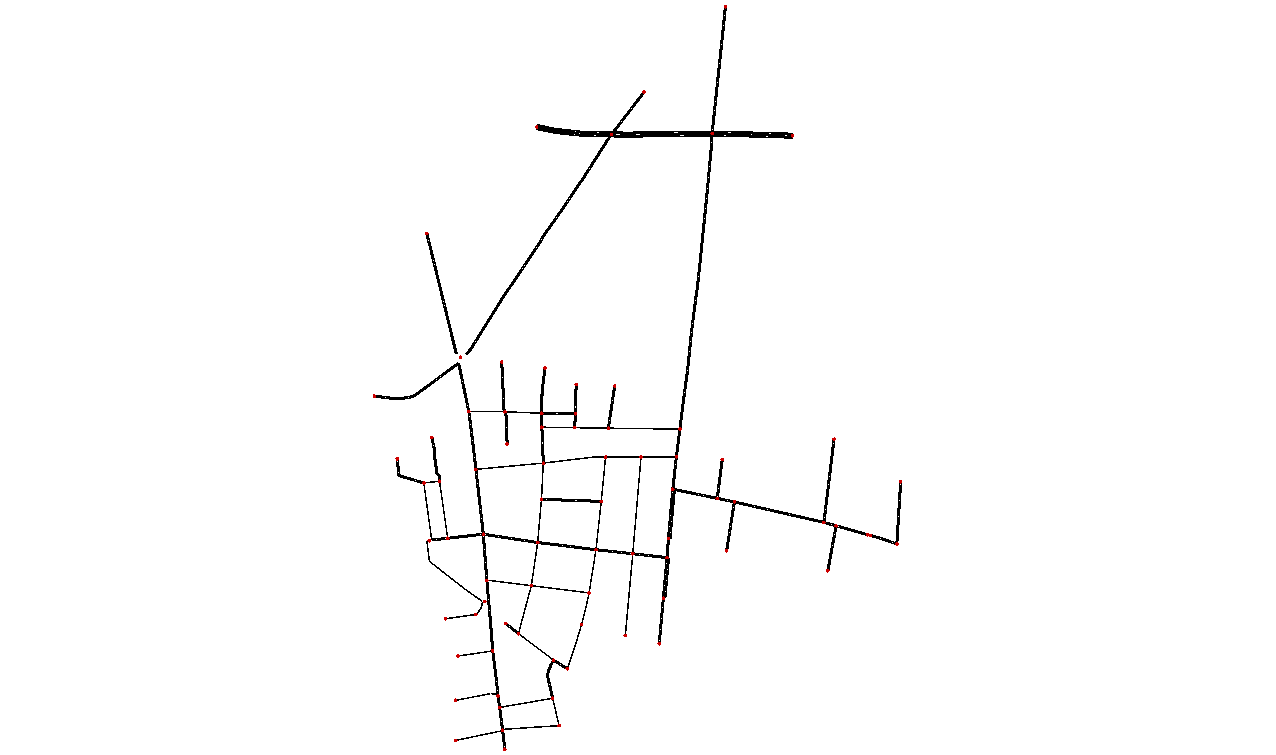}
        \caption{Part of Cologne network.}
        \label{fig:cologne_network}
    \end{subfigure}
    \hfill
    \begin{subfigure}[b]{0.45\textwidth}
        \centering
        \includegraphics[width=\textwidth]{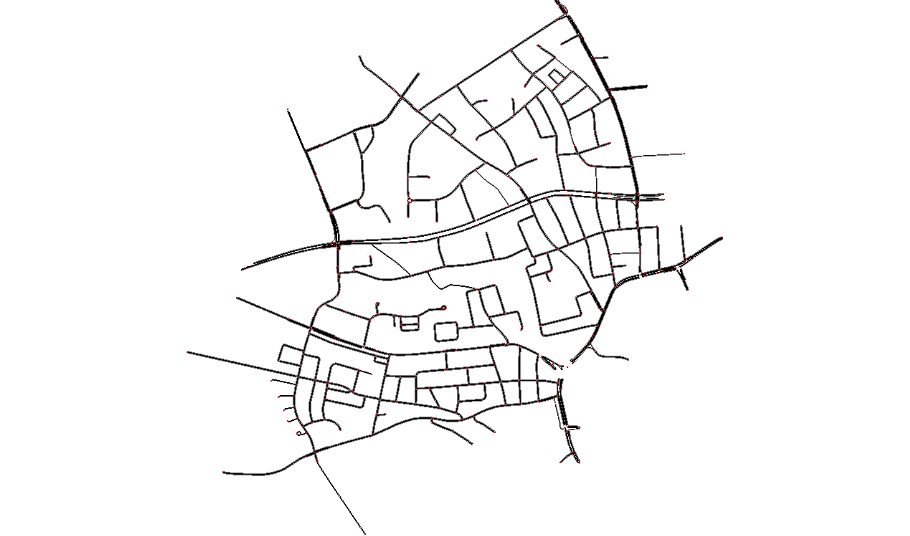}
        \caption{Part of Ingolstadt network.}
        \label{fig:ingolstadt_network}
    \end{subfigure}
    \hfill
    \begin{subfigure}[b]{0.45\textwidth}
        \centering
        \includegraphics[width=0.7\textwidth]{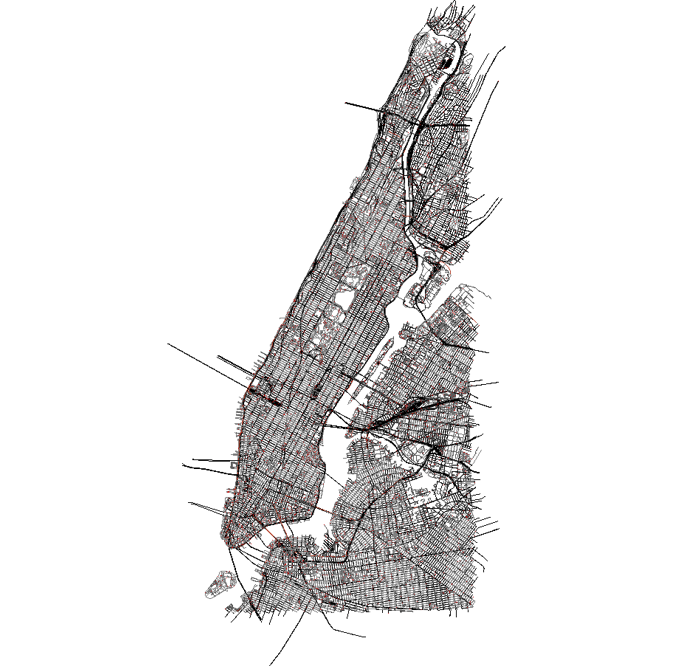}
        \caption{Manhattan network.}
        \label{fig:manhattan_network}
    \end{subfigure}
    \begin{subfigure}[b]{0.45\textwidth}
        \centering
        \includegraphics[width=0.5\textwidth]{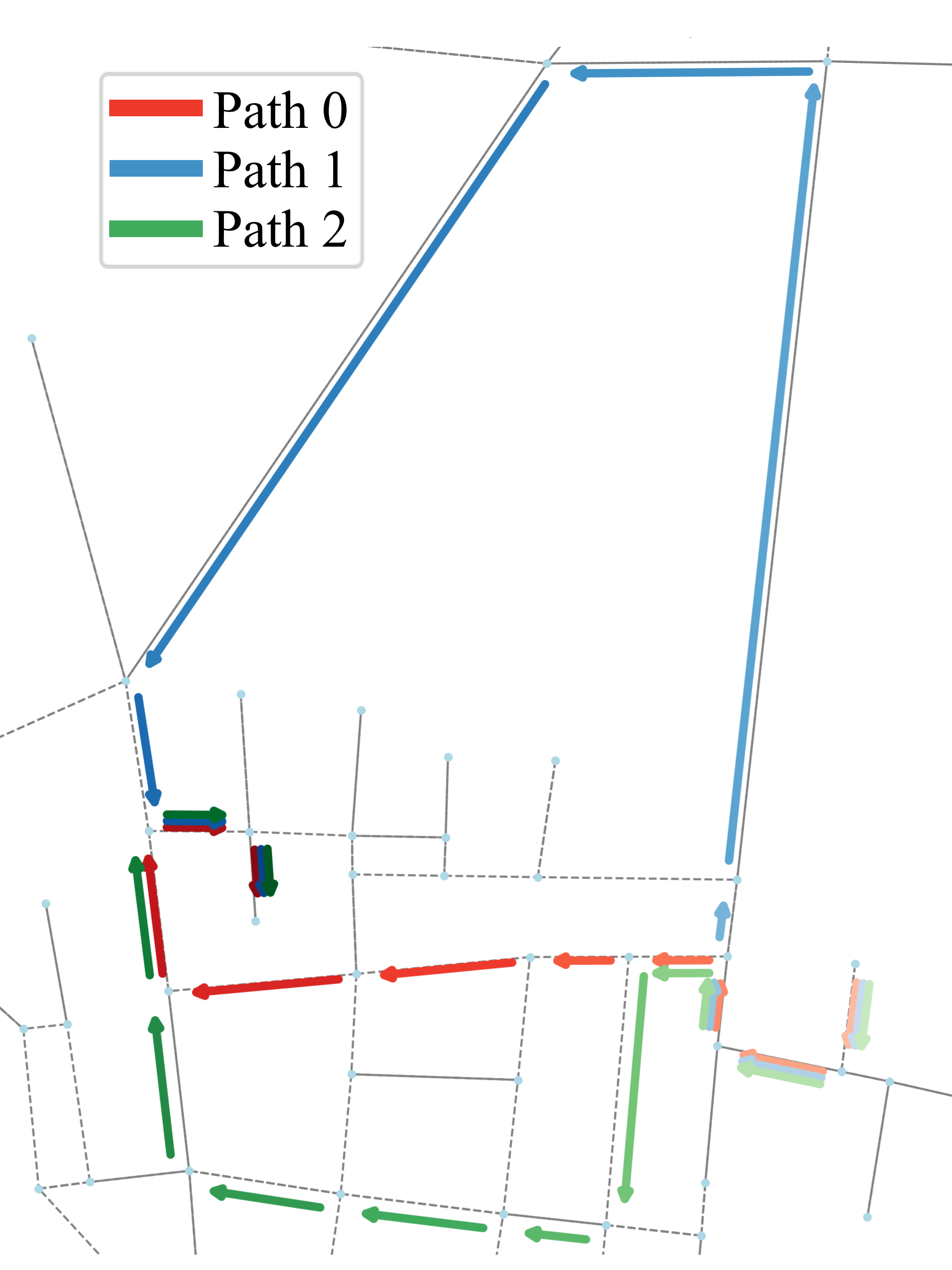}
        \caption{JanuX route generation in Cologne network.}
        \label{fig:routes}
    \end{subfigure}
    \caption{\textbf{Networks}: Small (part of Cologne), medium (part of Ingolstadt), and big (Manhattan) networks, which agents (human drivers and AVs) traverse to reach their destinations in illustrative experiments. Every day, agents choose routes from discrete options, illustrated on (d) for the Cologne network.}
    \label{fig:networks}
\end{figure}

RouteRL, thanks to the above functionalities, can be impactful for future scientific discoveries. We demonstrate it with three distinct traffic scenarios on the urban networks of Cologne, Ingolstadt, and Manhattan (Figure \ref{fig:networks}), which represent small, medium, and real-size problem instances respectively. As illustrated in Figure \ref{fig:routerl_flow} and the code snippet from the previous section, each experiment starts with only human agents, each learning the subjectively optimal route choices for 100 episodes (representing 100 days of commute). Every day agents choose from action space (routes between given origin-destination points), generated with JanuX, as illustrated in Figure \ref{fig:routes}. When human agents stabilize their choices (after 100 days), RouteRL is ready to introduce AV agents into the system. Typically, a predefined share of human drivers transition into AVs and start learning to find an efficient strategy using different MARL algorithms. After optimizing the policy to maximize expected rewards, the AVs enter the testing phase, where they execute 100 episodes to sample the performance of the learned policy.

We use the MARL algorithm implementations from TorchRL appropriate for respective experimental settings. For selfish AVs (Table \ref{tab:behaviors}), where agents are independent, we employ Independent Q-Learning (IQL) algorithm \cite{iql} as the initial baseline, and the Independent Proximal Policy Optimization (IPPO) algorithm \cite{ippo}, which proved efficiency in similar tasks \cite{yu2022surprisingeffectivenessppocooperative, papoudakis2021benchmarkingmultiagentdeepreinforcement}. For malicious strategies, where the reward signal is shared, we use algorithms designed for collaborative tasks: Value Decomposition Networks (VDN) \cite{vdn} (factorizes the joint Q-function as the sum of individual Q-values) and QMIX (adopts a mixing network with a monotonicity constraint) \cite{qmix}. Additionally, we report experiments with the multi-agent version of the Proximal Policy Optimization (PPO) algorithm \cite{ppo}, Multi-Agent PPO (MAPPO) \cite{mappo}.
Presented experiments, along with additional ones, are reproducible and described as tutorials on our repository, showcasing various experiment setups using RouteRL.

\begin{figure}[ht]
    \centering
    \includegraphics[width=\textwidth]{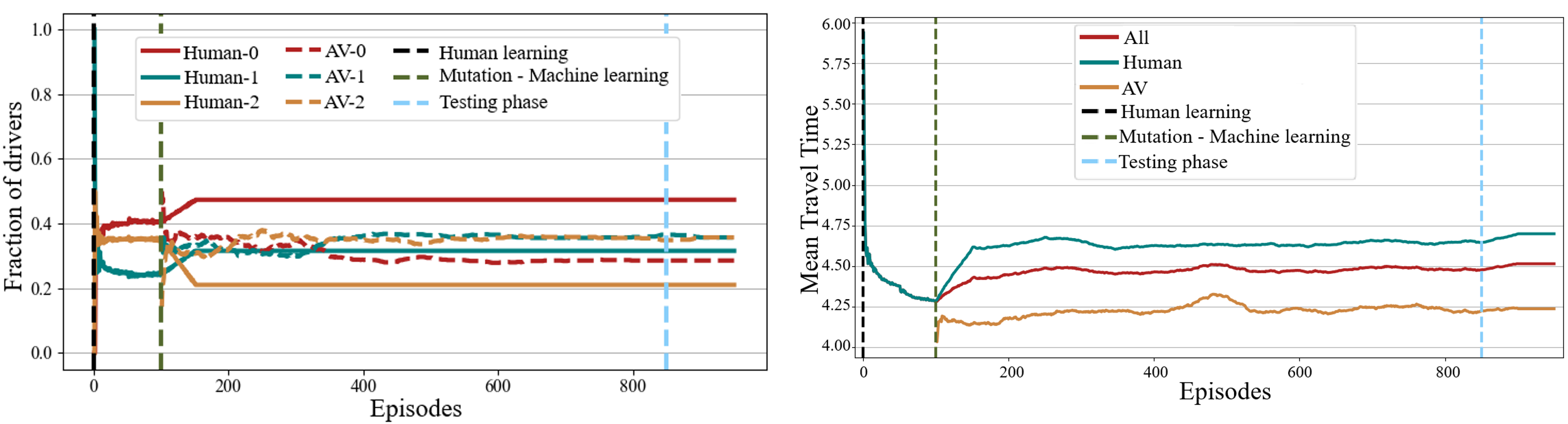}
    \caption{\textbf{Results from the experiment on the Cologne network.} 40 out of 100 human drivers transition into AVs with `malicious` behavior. 
    These results suggest that actions made by AVs in the Cologne network can be different from those made by humans (left) and AV's actions will not only make AVs arrive faster but also make the humans travel longer (right).}
    \label{fig:action_shifts_travel_times}
\end{figure}

\paragraph{Will AVs arrive earlier than human drivers in Cologne?}
Results from RouteRL experiment on our Cologne network reveal that the introduction of AVs into urban traffic can influence human agents' decision-making and increase their travel time. 
Figure \ref{fig:action_shifts_travel_times} illustrates the fraction of drivers split among available routes and the average travel time of all the agents in the system. Such insight highlights the need for research towards a better understanding of the effects of AV introduction, such as increased congestion or $\text{CO}_\text{2}$ emissions.

\begin{figure}[ht]
    \centering
    \includegraphics[width=\textwidth]{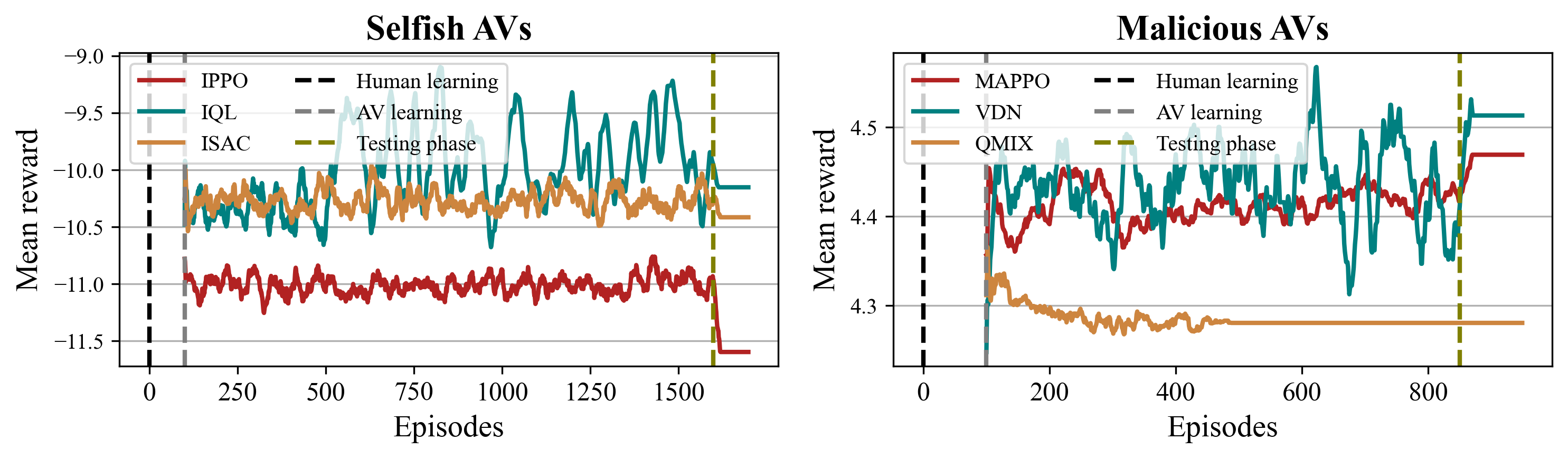}
    \caption{\textbf{Mean rewards of AVs  with various algorithms and strategies.} In Ingolstadt, 20 out of 100 human agents transition to selfish AVs (left), and in Cologne, 40 out of 100 human agents transition to malicious AVs (right). The results show IQL and VDN yield superior returns in their respective experiments.
}
    \label{fig:mean_rewards}
\end{figure}

\begin{figure}[ht]
    \centering
    \includegraphics[width=\textwidth]{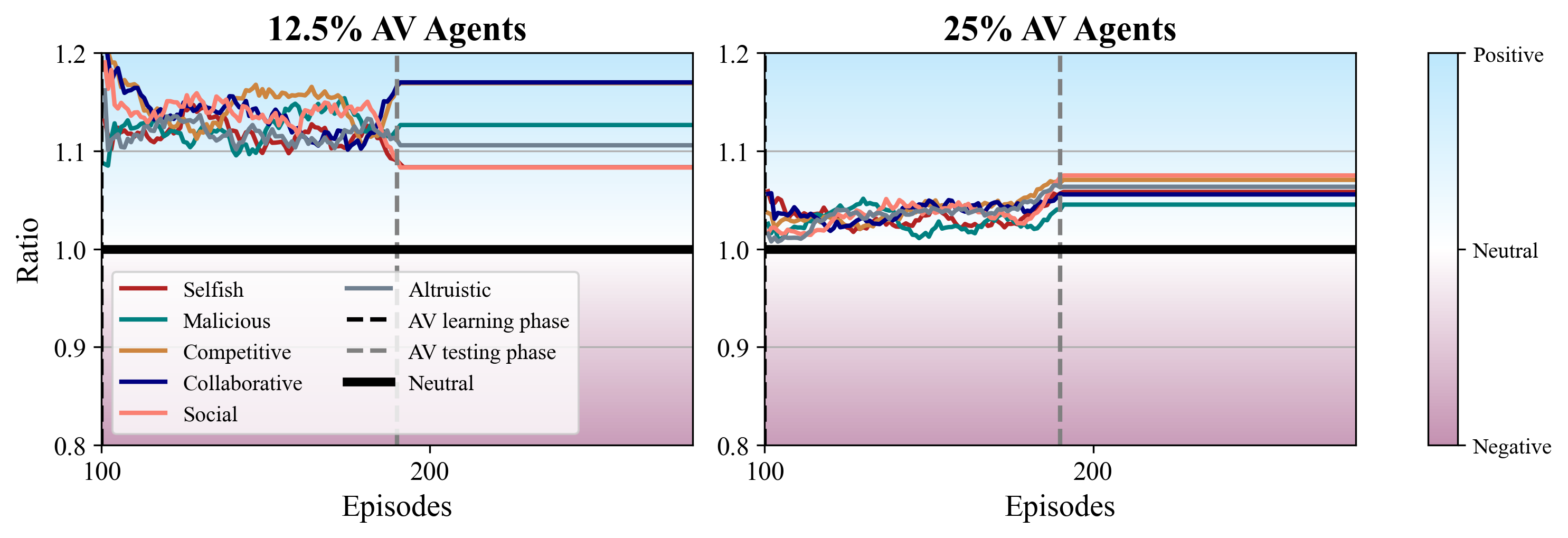}
    \caption{
    \textbf{AV advantage:} In the Manhattan network (Figure \ref{fig:manhattan_network}), being an AV provides an advantage regardless of the reward formulation. However, this advantage decreases with larger market shares. Y-axis is the ratio of the mean travel time of human drivers to that of AV agents, values greater than 1 indicate that being an AV is more advantageous than being a human driver.
    }
    \label{fig:cav_advantage}
\end{figure}

\paragraph{Which RL algorithms are efficient for collaborative and individual routing problems?}
AV agents can use different MARL algorithms to optimize their choices, each stabilizing in different solutions (Figure \ref{fig:mean_rewards}). With RouteRL, one can compare a variety of state-of-the-art MARL algorithms and benchmark them. Figure \ref{fig:mean_rewards} shows that while IQL worked best for individual reward, VDN performed best in the collaborative task. 

\paragraph{Will the benefit of AVs over humans increase with their growing share?}
Figure \ref{fig:cav_advantage} shows the ratio of average travel times of humans over AVs in our Manhattan experiment. 
In all tested AV behaviors, being an AV allows arriving earlier than human drivers (values greater than one). However, this benefit decreases when the share of AVs increases from 12.5\% to 20\%.

The sample results presented in this section highlight the significance, flexibility, and applicability of the RouteRL framework. RouteRL supports experiments with varying numbers of agents, urban networks, reward formulations, and share of AVs.

    \section{Impact}
\label{sec:impact}

AVs will soon enter our cities and start making routing decisions. Inevitably, some kind of algorithms will be used by commercial fleet operators (like Tesla or Waymo) to select, for their clients, optimal routes to get to their destinations. This is a complex optimization problem, hardly solvable with analytical methods for real-world scenarios where thousands of agents simultaneously select among hundreds of available routes in the networks with tens of thousands of links. Classic discrete optimization techniques fail, opening the way to machine learning alternatives, and as a natural candidate for dynamic systems, reinforcement learning.


RouteRL facilitates the modeling and prototyping of the collective route choices of individual AVs and fleets of AVs, by integrating human behavioral models and MARL algorithms into an agent-based traffic simulation. It supports a broad range of experimental schemes, where drivers transition into AVs, which then behave selfishly, collaboratively, or even maliciously. 
Its outputs are aimed to reveal the diverse socioethical aspects of this transition. By reporting various KPIs such as travel times, rewards, and agent preference changes, RouteRL reports the impact on congestion, travel times, and mileage, along with potentially undesirable social consequences. 
RouteRL scales to experiments on real-size networks with hundreds of agents. Its flexible parameterization, standard input/output formats, and integration of popular frameworks within its workflow allow researchers and system designers to pursue realistic and significant research questions, and ultimately, derive meaningful results and actionable insights.

The introduction of AVs into our cities opens new research questions that shall be answered upfront before commercial providers introduce black-box solutions without social control and participation. With the modeling framework of RouteRL we can, to some extent, answer experimentally the research questions such as (adapted from \cite{jamroz}):
\begin{compactitem}
    \item Will routing decisions of AVs differ substantially from choices made by human drivers?
    \item Will the impact on travel times depend on the AV strategy?
    \item Can a large fleet of AVs improve travel times for all the drivers?
    \item Does the benefits of AVs come at a price of equity?
    \item Are human populations prone to exploitation by collective AV fleets?
    \item Which MARL algorithms reliably solve AV routing problems?
    \item Can AV fleets contribute to policy goals (like $CO_2$ emissions or sustainability)?
    \item Can malicious AV strategies be detected?
    
\end{compactitem}

The community lacks tools to answer questions like the above reliably, and
RouteRL contributes with an evaluation ground, allowing more comprehensive assessments by incorporating the aforementioned complexities of the multi-agent route choice problem  
    for assessment, development, and comparative analyses.


While the day-to-day routing problem for AV fleets sharing road networks with human drivers can be formulated as a multi-agent decision-making process and thus solved with MARL, standard solutions are missing. The scientific community lacks a framework on which the algorithms can be tested, developed, and compared. The proper testbed must be interdisciplinary and realistically represent the road networks, travel demand patterns, traffic flow dynamics, and system performance. The performance needs to include views of individual agents (humans, whether driving or being driven in AVs), fleet operators, traffic managers, and policymakers. The desired algorithmic solutions shall, apart from being computationally efficient, be socially responsible, balancing often opposing objectives of various parties involved in the complex social system of urban traffic. RouteRL is designed to serve as a testing ground for new, hopefully, efficient and reliable solutions.



The multi-agent route choice problem poses a significant algorithmic challenge due to its multifaceted and dynamic nature. It is interdisciplinary and requires expertise in traffic engineering, microeconomics of discrete choices, urban policies, optimization, and machine learning. It suffers from the curse of dimensionality, the environment is at most partially observable and 
non-stationary, which requires adaptive solutions. Maximizing the fleet's shared reward requires communication protocols and coordination mechanisms which, for real-world AV scenarios with high-dimensional state-action spaces, are difficult to scale. Finally, the presence of human drivers,
 who are bounded-rational decision-makers with heterogenous and non-deterministic route-choice behavior, makes the setting non-deterministic, specifically if we include human adaptation to AV actions. 
Altogether this makes the problem significantly harder than standard MARL benchmarks (like \cite{pettingzoo, jaxmarl, nmmo, ma_comp}) and calls for a unified development framework like RouteRL.

    \section{Conclusions}
\label{sec:conclusions}

In this paper, we introduced \textbf{RouteRL}, a MARL framework for multi-agent urban route choice. RouteRL is a python open-source package aimed to enhance our understanding of future urban mobility with AVs. It supports various experimental settings through configurable traffic networks, human behavioral models, AV market configuration, and AV algorithms. With its modular architecture and integration with standardized frameworks, RouteRL offers reusability and simplifies experiment setup. Moreover, full integration with the TorchRL library allows the development and testing of state-of-the-art MARL algorithms in solving the route choice problem. By providing analytical insights from experiments, RouteRL enables researchers to systematically underpin their findings and explore alternative strategies for more efficient future traffic systems.
We believe that a class of open research problems arising around the routing behavior of AVs in future cities can be answered with the help of RouteRL.

To ensure accessibility and alignment with community standards, we provide RouteRL with comprehensive documentation, release it under an open-source license, host it in a public repository and, for faster adoption, we provide a reusable code capsule with a reproducible experiment.

\section*{Acknowledgements}
\label{sec:acknowledgements}
This research is financed by the European Union within the Horizon Europe Framework Programme, ERC Starting Grant number 101075838: COeXISTENCE.

\section*{Declaration of generative AI and AI-assisted technologies in the writing process}

During the preparation of this work, the authors used ChatGPT (GPT-4o) for proofreading and minor style corrections. After using this tool/service, the authors reviewed and edited the content as needed and take full responsibility for the content of the published article.
    
    \bibliographystyle{elsarticle-num} 
    \bibliography{references.bib}

\begin{thebibliography}{10}
\expandafter\ifx\csname url\endcsname\relax
  \def\url#1{\texttt{#1}}\fi
\expandafter\ifx\csname urlprefix\endcsname\relax\def\urlprefix{URL }\fi
\expandafter\ifx\csname href\endcsname\relax
  \def\href#1#2{#2} \def\path#1{#1}\fi

\bibitem{av_predictions}
T.~Litman, Autonomous vehicle implementation predictions: Implications for transport planning, Victoria Transport Policy Institute (2020).

\bibitem{urban_future}
M.~Miskolczi, D.~Földes, A.~Munkácsy, M.~Jászberényi, Urban mobility scenarios until the 2030s, Sustainable Cities and Society 72 (2021) 103029.
\newblock \href {https://doi.org/https://doi.org/10.1016/j.scs.2021.103029} {\path{doi:https://doi.org/10.1016/j.scs.2021.103029}}.

\bibitem{SUMO}
P.~A. Lopez, M.~Behrisch, L.~Bieker-Walz, J.~Erdmann, Y.-P. Fl{\"o}tter{\"o}d, R.~Hilbrich, L.~L{\"u}cken, J.~Rummel, P.~Wagner, E.~Wie{\ss}ner, \href{https://elib.dlr.de/124092/}{Microscopic traffic simulation using sumo}, in: The 21st IEEE International Conference on Intelligent Transportation Systems, IEEE, 2018.
\newline\urlprefix\url{https://elib.dlr.de/124092/}

\bibitem{OpenStreetMap}
{OpenStreetMap contributors}, {Planet dump retrieved from https://planet.osm.org }, \url{ https://www.openstreetmap.org } (2017).

\bibitem{wardrop_road_1952}
J.~G. Wardrop, Road paper. some theoretical aspects of road traffic research., Proceedings of the institution of civil engineers 1~(3) (1952) 325--362, publisher: Thomas Telford-ICE Virtual Library.

\bibitem{torchrl}
A.~Bou, M.~Bettini, S.~Dittert, V.~Kumar, S.~Sodhani, X.~Yang, G.~D. Fabritiis, V.~Moens, \href{https://arxiv.org/abs/2306.00577}{Torchrl: A data-driven decision-making library for pytorch} (2023).
\newblock \href {http://arxiv.org/abs/2306.00577} {\path{arXiv:2306.00577}}.
\newline\urlprefix\url{https://arxiv.org/abs/2306.00577}

\bibitem{qmix}
T.~Rashid, M.~Samvelyan, C.~S. De~Witt, G.~Farquhar, J.~Foerster, S.~Whiteson, Monotonic value function factorisation for deep multi-agent reinforcement learning, Journal of Machine Learning Research 21~(178) (2020) 1--51.

\bibitem{Treiber_2000}
M.~Treiber, A.~Hennecke, D.~Helbing, \href{http://dx.doi.org/10.1103/PhysRevE.62.1805}{Congested traffic states in empirical observations and microscopic simulations}, Physical Review E 62~(2) (2000).
\newblock \href {https://doi.org/10.1103/physreve.62.1805} {\path{doi:10.1103/physreve.62.1805}}.
\newline\urlprefix\url{http://dx.doi.org/10.1103/PhysRevE.62.1805}

\bibitem{janux}
A.~O. Akman, \href{https://github.com/COeXISTENCE-PROJECT/JanuX}{{JanuX}} (Feb. 2025).
\newline\urlprefix\url{https://github.com/COeXISTENCE-PROJECT/JanuX}

\bibitem{akman_impact}
A.~O. Akman, A.~Psarou, Z.~G. Varga, G.~Jamr{\'o}z, R.~Kucharski, \href{https://openreview.net/forum?id=88zP8xh5D2}{Impact of collective behaviors of autonomous vehicles on urban traffic dynamics: A multi-agent reinforcement learning approach}, in: Seventeenth European Workshop on Reinforcement Learning, 2024.
\newline\urlprefix\url{https://openreview.net/forum?id=88zP8xh5D2}

\bibitem{jamroz}
G.~Jamr{\'o}z, A.~O. Akman, A.~Psarou, Z.~G. Varga, R.~Kucharski, \href{https://arxiv.org/pdf/2409.12839}{Social impact of cavs--coexistence of machines and humans in the context of route choice}, Scientific Reports, in press (2025).
\newline\urlprefix\url{https://arxiv.org/pdf/2409.12839}

\bibitem{gawron}
C.~Gawron, Simulation-based traffic assignment, Ph.D. thesis, Universitat zu Koln (1998).

\bibitem{culo}
J.~Li, Z.~Wang, Y.~Nie, Wardrop equilibrium can be boundedly rational: A new behavioral theory of route choice, Transportation Science 58~(5) (2024) 973--994.

\bibitem{cascetta}
E.~Cascetta, Transportation systems analysis: models and applications, Vol.~29, Springer Science \& Business Media, 2009.

\bibitem{resco}
J.~Ault, G.~Sharon, \href{https://openreview.net/forum?id=LqRSh6V0vR}{Reinforcement learning benchmarks for traffic signal control}, in: Proceedings of the Thirty-fifth Conference on Neural Information Processing Systems (NeurIPS 2021) Datasets and Benchmarks Track, 2021.
\newline\urlprefix\url{https://openreview.net/forum?id=LqRSh6V0vR}

\bibitem{sumo_repo}
Eclipse, \href{https://github.com/eclipse-sumo/sumo}{sumo} (Feb. 2025).
\newline\urlprefix\url{https://github.com/eclipse-sumo/sumo}

\bibitem{sumorl}
L.~N. Alegre, {SUMO-RL}, \url{https://github.com/LucasAlegre/sumo-rl} (2019).

\bibitem{iql}
M.~Tan, Multi-agent reinforcement learning: independent vs. cooperative agents, Morgan Kaufmann Publishers Inc., San Francisco, CA, USA, 1997, p. 487–494.

\bibitem{ippo}
C.~S. de~Witt, T.~Gupta, D.~Makoviichuk, V.~Makoviychuk, P.~H.~S. Torr, M.~Sun, S.~Whiteson, \href{https://arxiv.org/abs/2011.09533}{Is independent learning all you need in the starcraft multi-agent challenge?}, CoRR abs/2011.09533 (2020).
\newblock \href {http://arxiv.org/abs/2011.09533} {\path{arXiv:2011.09533}}.
\newline\urlprefix\url{https://arxiv.org/abs/2011.09533}

\bibitem{yu2022surprisingeffectivenessppocooperative}
C.~Yu, A.~Velu, E.~Vinitsky, J.~Gao, Y.~Wang, A.~Bayen, Y.~Wu, The surprising effectiveness of ppo in cooperative multi-agent games, Advances in Neural Information Processing Systems 35 (2022) 24611--24624.

\bibitem{papoudakis2021benchmarkingmultiagentdeepreinforcement}
G.~Papoudakis, F.~Christianos, L.~Sch{\"a}fer, S.~V. Albrecht, \href{https://api.semanticscholar.org/CorpusID:235417602}{Benchmarking multi-agent deep reinforcement learning algorithms in cooperative tasks}, in: NeurIPS Datasets and Benchmarks, 2020.
\newline\urlprefix\url{https://api.semanticscholar.org/CorpusID:235417602}

\bibitem{vdn}
P.~Sunehag, G.~Lever, A.~Gruslys, W.~M. Czarnecki, V.~Zambaldi, M.~Jaderberg, M.~Lanctot, N.~Sonnerat, J.~Z. Leibo, K.~Tuyls, T.~Graepel, Value-decomposition networks for cooperative multi-agent learning based on team reward, in: Proceedings of the 17th International Conference on Autonomous Agents and MultiAgent Systems, AAMAS '18, International Foundation for Autonomous Agents and Multiagent Systems, Richland, SC, 2018, p. 2085–2087.

\bibitem{ppo}
J.~Schulman, F.~Wolski, P.~Dhariwal, A.~Radford, O.~Klimov, \href{https://arxiv.org/abs/1707.06347}{Proximal policy optimization algorithms} (2017).
\newblock \href {http://arxiv.org/abs/1707.06347} {\path{arXiv:1707.06347}}.
\newline\urlprefix\url{https://arxiv.org/abs/1707.06347}

\bibitem{mappo}
C.~Yu, A.~Velu, E.~Vinitsky, J.~Gao, Y.~Wang, A.~Bayen, Y.~WU, \href{https://proceedings.neurips.cc/paper_files/paper/2022/file/9c1535a02f0ce079433344e14d910597-Paper-Datasets_and_Benchmarks.pdf}{The surprising effectiveness of ppo in cooperative multi-agent games}, in: Advances in Neural Information Processing Systems, Vol.~35, Curran Associates, Inc., 2022, pp. 24611--24624.
\newline\urlprefix\url{https://proceedings.neurips.cc/paper_files/paper/2022/file/9c1535a02f0ce079433344e14d910597-Paper-Datasets_and_Benchmarks.pdf}

\bibitem{pettingzoo}
J.~Terry, B.~Black, N.~Grammel, M.~Jayakumar, A.~Hari, R.~Sullivan, L.~S. Santos, C.~Dieffendahl, C.~Horsch, R.~Perez-Vicente, N.~Williams, Y.~Lokesh, P.~Ravi, \href{https://proceedings.neurips.cc/paper_files/paper/2021/file/7ed2d3454c5eea71148b11d0c25104ff-Paper.pdf}{Pettingzoo: Gym for multi-agent reinforcement learning}, in: Advances in Neural Information Processing Systems, Vol.~34, Curran Associates, Inc., 2021, pp. 15032--15043.
\newline\urlprefix\url{https://proceedings.neurips.cc/paper_files/paper/2021/file/7ed2d3454c5eea71148b11d0c25104ff-Paper.pdf}

\bibitem{jaxmarl}
A.~Rutherford, B.~Ellis, M.~Gallici, J.~Cook, A.~Lupu, G.~Ingvarsson, T.~Willi, R.~Hammond, A.~Khan, C.~S. de~Witt, A.~Souly, S.~Bandyopadhyay, M.~Samvelyan, M.~Jiang, R.~T. Lange, S.~Whiteson, B.~Lacerda, N.~Hawes, T.~Rocktaschel, C.~Lu, J.~N. Foerster, \href{https://arxiv.org/abs/2311.10090}{Jaxmarl: Multi-agent rl environments and algorithms in jax} (2024).
\newblock \href {http://arxiv.org/abs/2311.10090} {\path{arXiv:2311.10090}}.
\newline\urlprefix\url{https://arxiv.org/abs/2311.10090}

\bibitem{nmmo}
J.~Suarez, Y.~Du, I.~Mordach, P.~Isola, Neural mmo v1.3: A massively multiagent game environment for training and evaluating neural networks, in: Proceedings of the 19th International Conference on Autonomous Agents and MultiAgent Systems, AAMAS '20, International Foundation for Autonomous Agents and Multiagent Systems, Richland, SC, 2020, p. 2020–2022.

\bibitem{ma_comp}
T.~Bansal, J.~Pachocki, S.~Sidor, I.~Sutskever, I.~Mordatch, \href{https://arxiv.org/abs/1710.03748}{Emergent complexity via multi-agent competition} (2017).
\newblock \href {http://arxiv.org/abs/1710.03748} {\path{arXiv:1710.03748}}.
\newline\urlprefix\url{https://arxiv.org/abs/1710.03748}

\end{thebibliography}
    
    \newpage
\appendix

\section{UML Class Diagram}
\label{sec:class_diagram}

    \begin{figure}[!ht]
        \centering
        \includegraphics[width=0.99\textwidth]{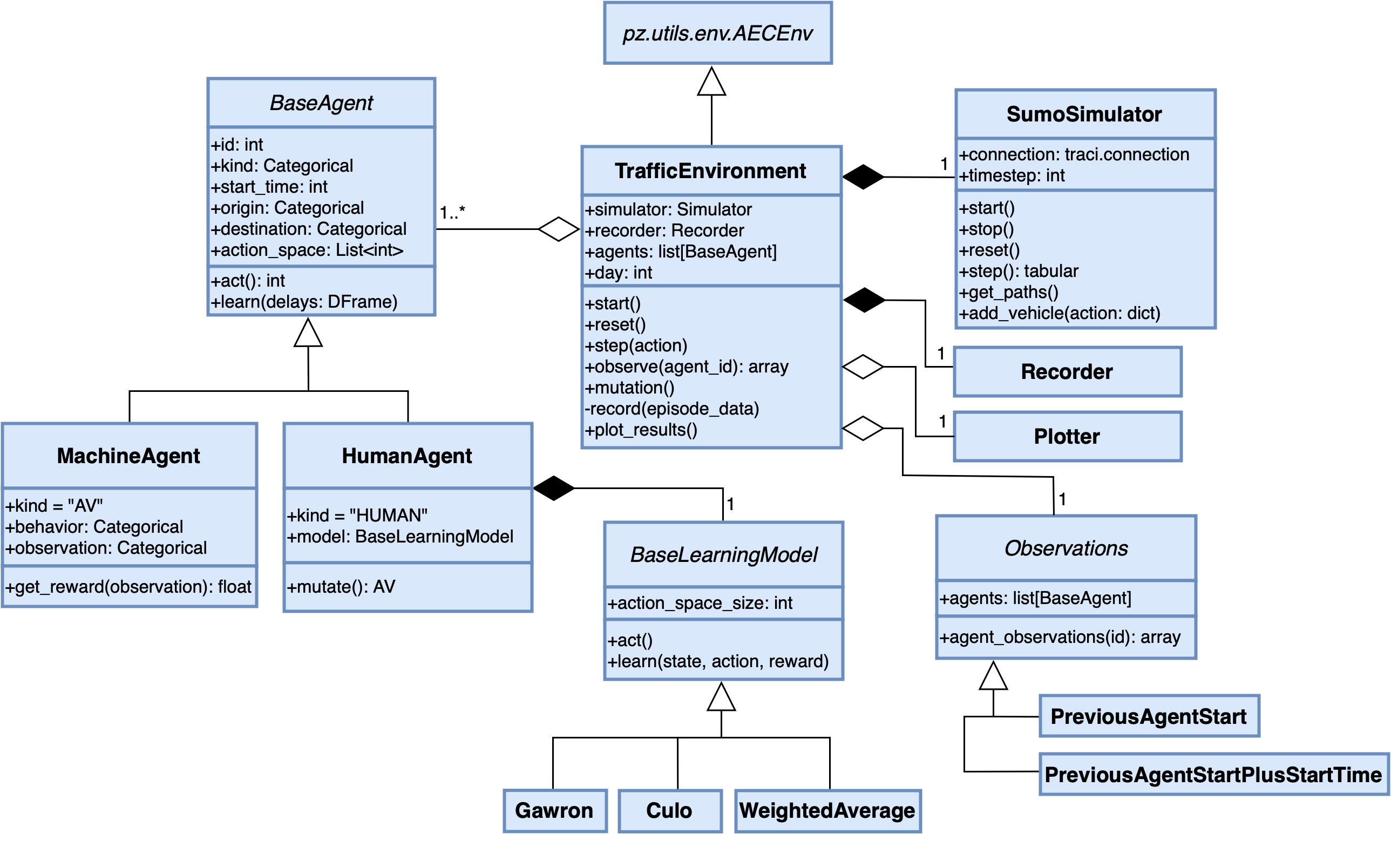}
        \caption{
        A high-level UML class diagram visualizing the class hierarchy within RouteRL. Inessential attributes and methods are omitted.
        }
        \label{fig:class_diagram}
    \end{figure}

\section{Sequence Diagram}
\label{sec:seq_diagram}

    \begin{figure}[H]
        \centering
        \includegraphics[height=0.83\textheight, width=\textwidth, keepaspectratio]{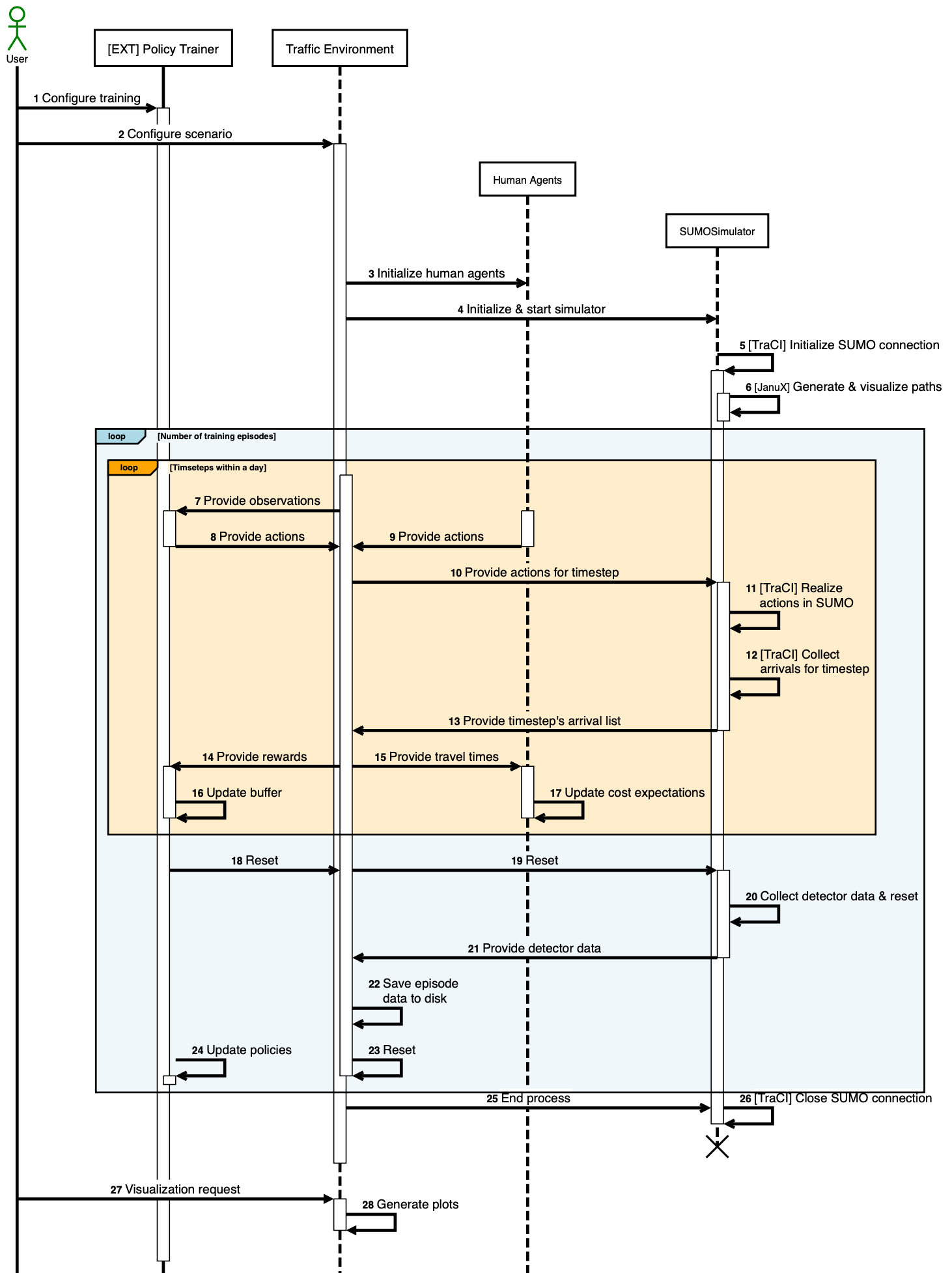}
        \caption{
            A UML sequence diagram depicting the user's interaction with different elements of RouteRL. External dependencies are provided in square brackets. Some intermediate steps and inessential modules are abstracted for clarity. Multiple instances (like a group of agents) are summarized into one participant.
        }
        \label{fig:sequence_diagram}
    \end{figure}

\section{Plotter results}
\label{sec:plots}

\begin{figure}[!ht]
\centering
    \begin{subfigure}[b]{0.49\textwidth}
        \centering
        \includegraphics[width=\textwidth]{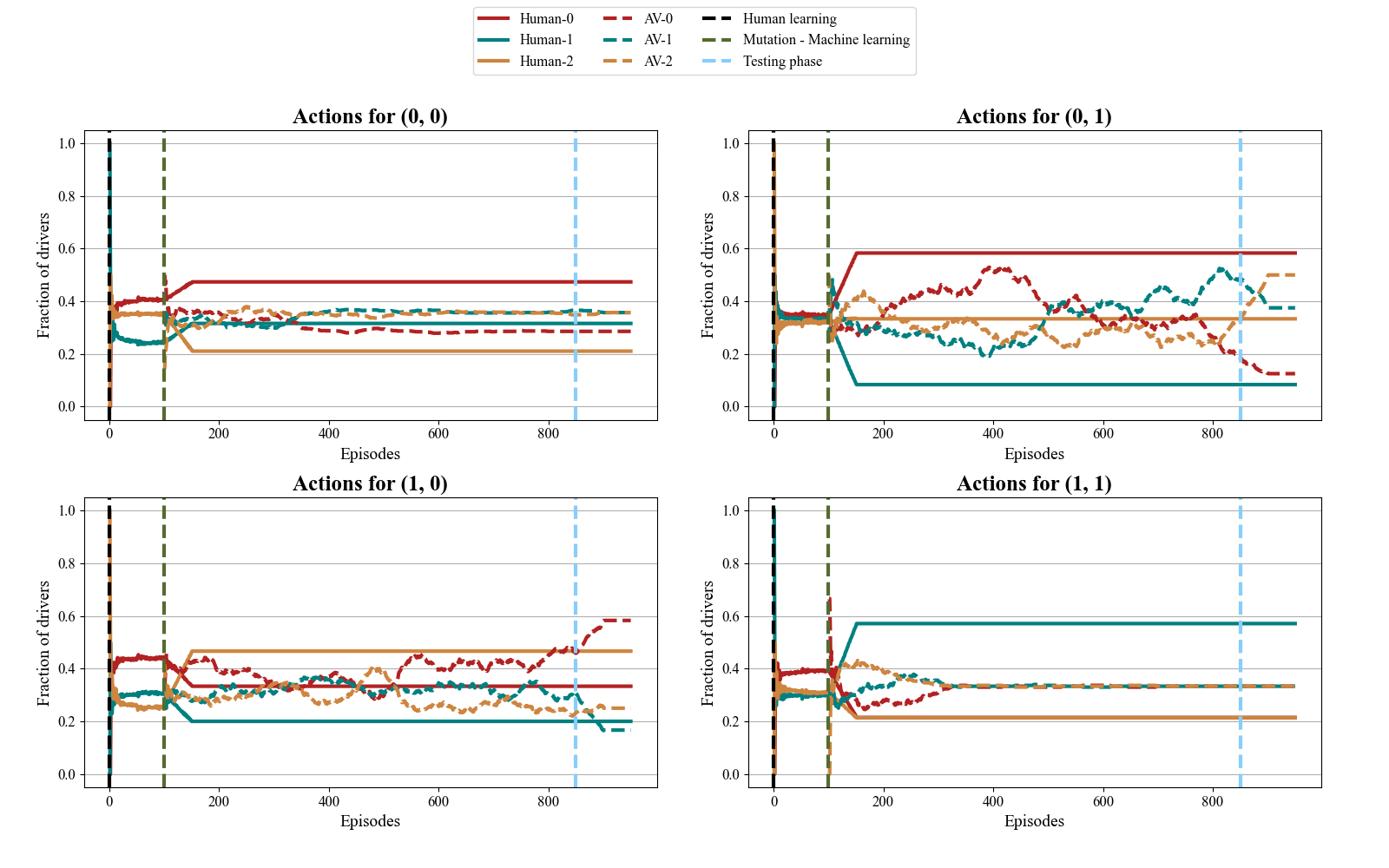}
        \caption{Fraction of humans and AV agents that choose one of the three available routes.}
        \label{fig:mappo_action_shifts}
    \end{subfigure}
    \hfill
    \begin{subfigure}[b]{0.49\textwidth}
        \centering
        \includegraphics[width=\textwidth]{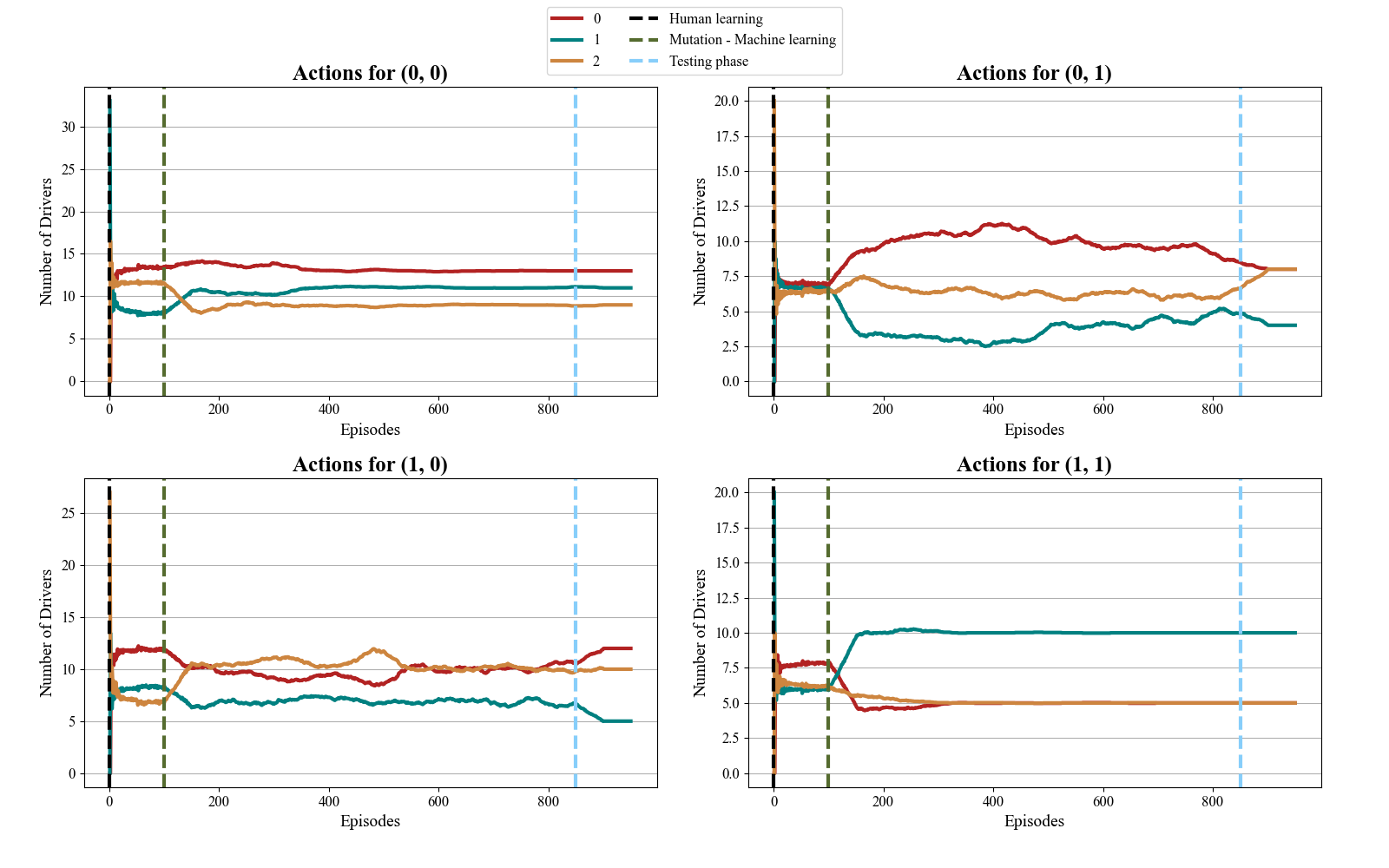}
        \caption{Fraction of drivers that choose one of the three available routes.}
        \label{fig:mappo_actions}
    \end{subfigure}
    \hfill
    \begin{subfigure}[b]{0.49\textwidth}
        \centering
        \includegraphics[width=\textwidth]{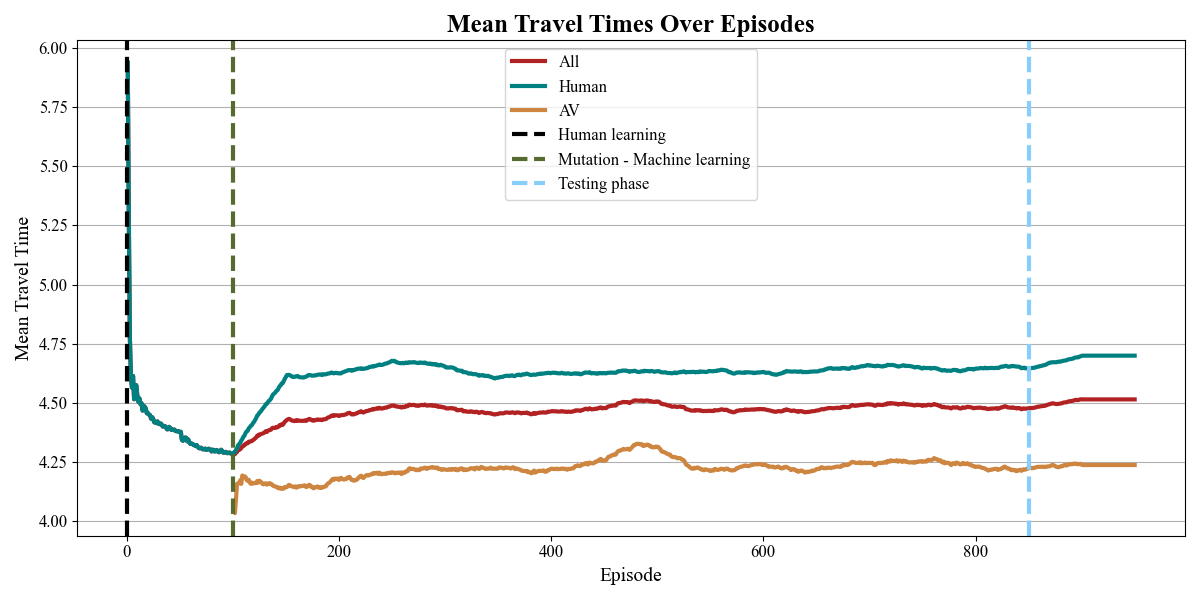}
        \caption{Mean travel times of all  agent groups present in the system.}
        \label{fig:mappo_travel_times}
    \end{subfigure}
    \hfill
    \begin{subfigure}[b]{0.49\textwidth}
        \centering
        \includegraphics[width=\textwidth]{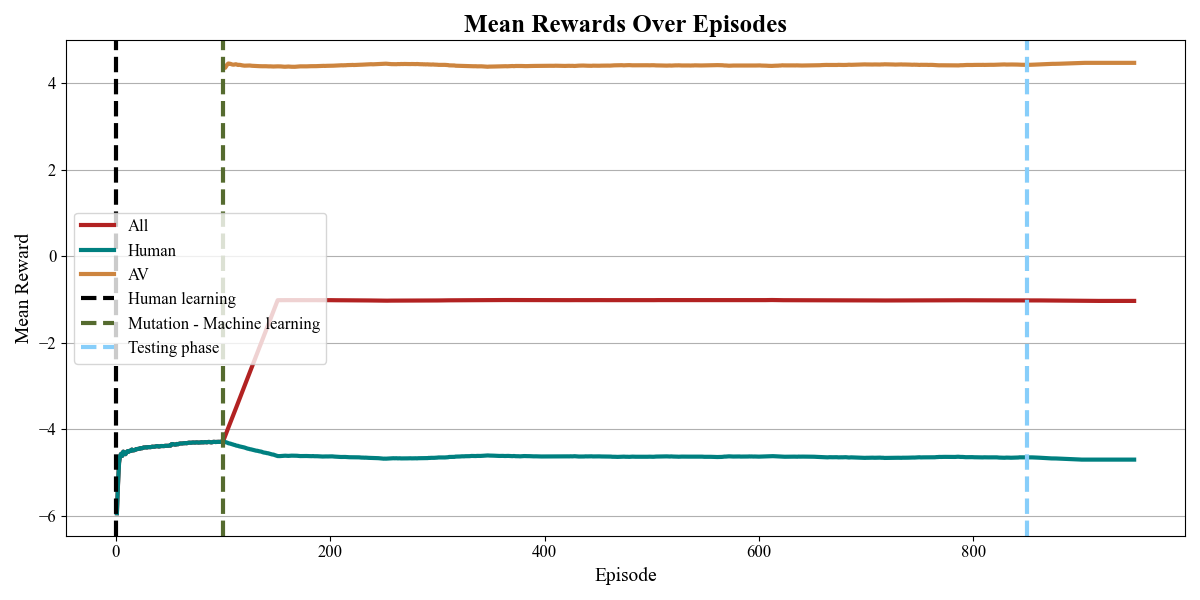}
        \caption{Mean reward of all agent groups present in the system.}
        \label{fig:mappo_rewards}
    \end{subfigure}
    \begin{subfigure}[b]{0.49\textwidth}
        \centering
        \includegraphics[width=\textwidth]{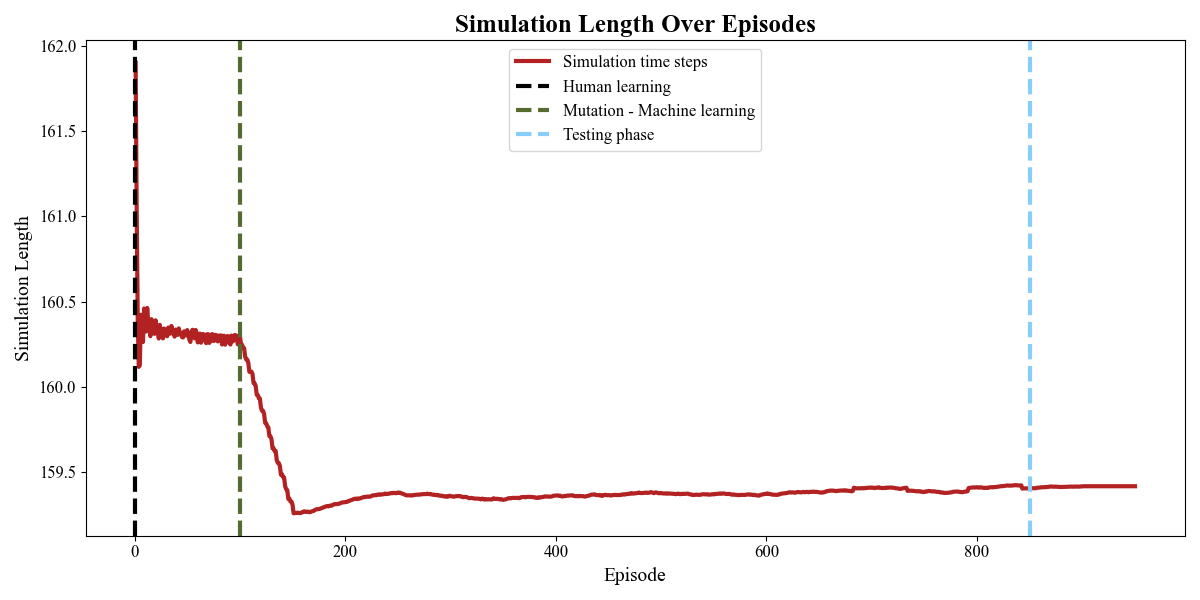}
        \caption{Simulation length over episodes.}
        \label{fig:mappo_simulation_length}
    \end{subfigure}
    \hfill
    \begin{subfigure}[b]{0.49\textwidth}
        \centering
        \includegraphics[width=\textwidth]{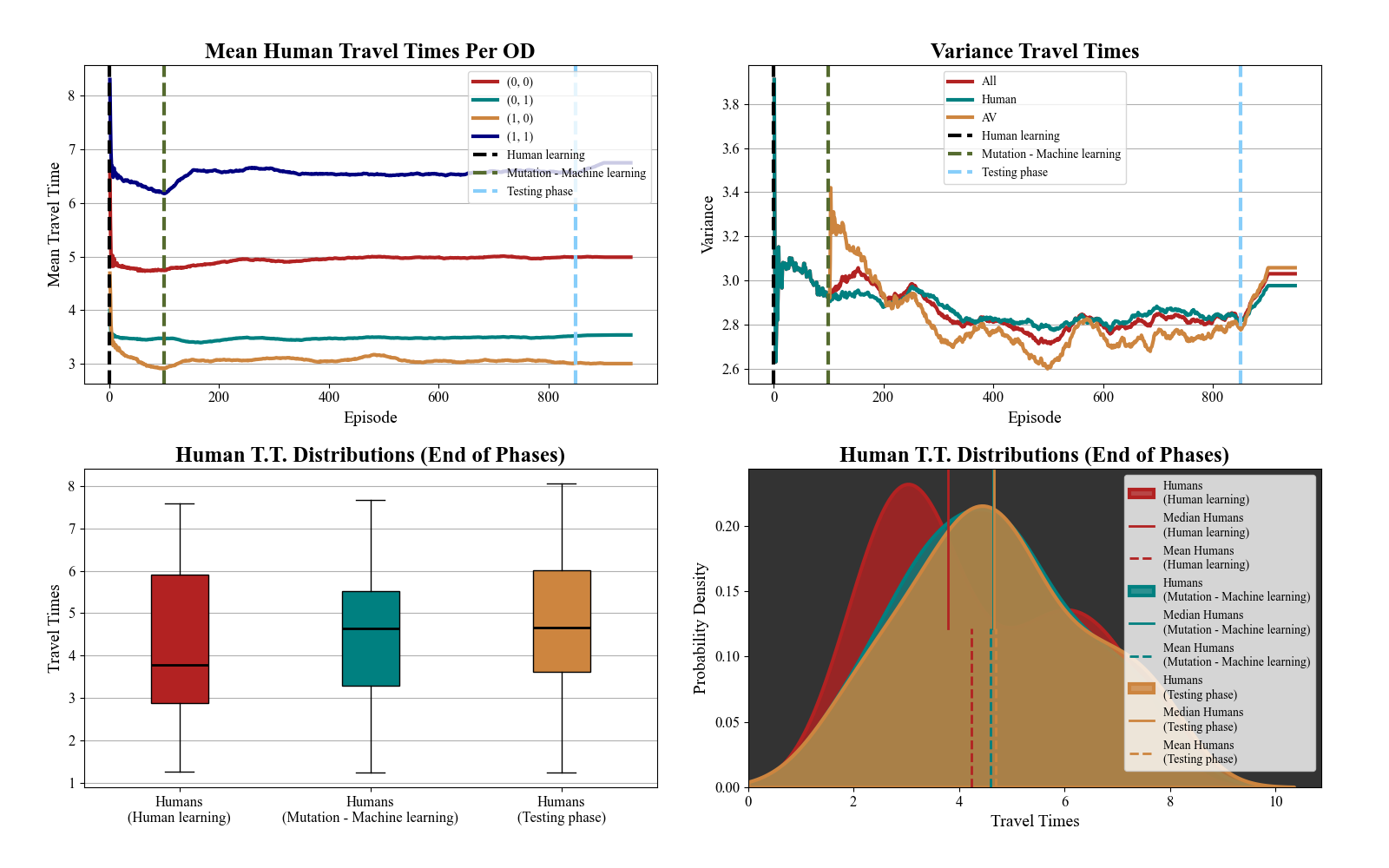}
        \caption{Travel time distribution plots.}
        \label{fig:mappo_tt_distributions}
    \end{subfigure}
    \caption{Plots generated by the 
    \texttt{Plotter} class of the RouteRL framework. This experiment is conducted in the Cologne network (Figure \ref{fig:cologne_network}). Initially, there are 100 human agents in the system, and 40 of these agents transition into AVs with malicious behavior, trained using the MAPPO algorithm.}
    \label{fig:plotter_results}
    \end{figure}
    
\end{document}